\documentclass[aps,prl,twocolumn,superscriptaddress,floats]{revtex4-1}
\usepackage{txfonts}
\usepackage{amssymb}
\usepackage{graphicx}
\begin{document} \hbadness=10000
\topmargin -0.8cm\oddsidemargin = -0.7cm\evensidemargin = -0.7cm
\title{Titanium trisulfide monolayer: A new direct-gap semiconductor with high and anisotropic carrier mobility}
\author{Jun Dai}
\affiliation{Department of Chemistry, University of Nebraska-Lincoln, Lincoln, NE 68588, USA}
\author{Xiao Cheng Zeng}
\email{xzeng1@unl.edu}
\affiliation{Department of Chemistry, University of Nebraska-Lincoln, Lincoln, NE 68588, USA}

\date{\today}
\begin{abstract}
A new two-dimensional (2D) layered material, namely, titanium trisulfide (TiS$_3$) monolayer sheet, is predicted to possess desired electronic properties for nanoelectronic applications. On basis of the first-principles calculations within the framework of density functional theory and deformation theory, we show that the TiS$_3$ 2D crystal is a direct gap semiconductor with a band gap of 1.06 eV and high carrier mobility. More remarkably, the in-plane electron mobility of the 2D  TiS$_3$ is highly anisotropic, amounting to $\sim$10,000 cm$^2$V$^{-1}$s$^{-1}$ in the \emph{b} direction, which is higher than that of the MoS$_2$ monolayer. Meanwhile, the hole mobility is about two orders of magnitude lower. We also find that bulk TiS$_3$ possesses lower cleavage energy than graphite, indicating high possibility of  exfoliation for TiS$_3$ monolayers or multilayers. Both dynamical and thermal stability of the  TiS$_3$ monolayer is examined via phonon-spectrum calculation and Born-Oppenheimer molecular dynamics simulation in \emph{NPT} ensemble. The predicted novel electronic properties render the TiS$_3$ monolayer an attractive 2D material for applications in future nanoelectronics.
\end{abstract}
\maketitle

The successful isolation of two-dimensional (2D) graphene in 2004 \cite{ref1} has captivated great research interests in 2D materials, particularly atomic-layered materials with weak interlayer van der Waals bonding\cite{ref2}. Besides the graphene\cite{ref3,ref4,ref5,ref6}, the family of 2D layered materials also include transition metal dichalcogenides (TMDCs)\cite{ref2,ref7,ref8}, hexagonal boron nitride (h-BN)\cite{ref9,ref10}, silicene\cite{ref11,ref12,ref13}, germanene\cite{ref14}, and phosphorene\cite{ref15,ref16}, among others. These 2D atomic-layered materials exhibit not only novel 2D geometries as they represent the thinnest crystalline solids that can be formed, but also new and exotic condensed matter phenomena that are absent in their bulk counterparts\cite{ref2,ref8}. For example, single graphene sheet is a zero-gap semiconductor with a linear Dirac-like dispersion near the Fermi level, while bulk graphite is known to exhibit semimetallic behavior with bandgap overlap of $\sim$41 meV.\cite{ref17} A MoS$_2$ monolayer sheet possesses a direct bandgap of $\sim$1.8 eV, while bulk MoS$_2$ possesses an indirect bandgap of 1.29 eV\cite{ref18}. The bandgap of a few layer phosphorene (a 2D form of black phosphorus) is highly layer-dependent. For the phosphorene monolayer, the bandgap is $\sim$1.5 eV while for the bulk phosphorus, the bandgap is merely $\sim$0.3 eV\cite{ref16,ref19,ref20,ref21,ref22}.

2D layered materials offer opportunities for a variety of applications, particularly in next-generation electronic devices such as field-effect transistors (FET) and logic circuits. For high-performance FET applications, a 2D material should possess a moderate bandgap and reasonably high in-plane carrier mobility. Graphene is a highly promising 2D material for high-speed nanotransistors due to its massless charge carriers. However, it lacks a bandgap for controllable operations.\cite{ref23,ref24,ref25} The molybdenum disulfide (MoS$_2$) monolayer sheets are more promising for FET applications since not only they possesses a direct bandgap of $\sim$1.8 eV,\cite{ref18} but also the 2D MoS$_2$-based FET devices show good performance with a high on/off ratio of $\sim$10$^8$ as well as a carrier mobility of $\sim$200 cm$^2$V$^{-1}$s$^{-1}$. The latter can be enhanced even up to 500 cm$^2$V$^{-1}$s$^{-1}$ with improvement.\cite{ref26,ref27} Also, recent experiments demonstrated that FET devices built upon few-layer phosphorene exhibit reasonably high on/off ratio (up to 10$^4$) and appreciably high hole mobility of $\sim$55 cm$^2$V$^{-1}$s$^{-1}$ (at a thickness of $\sim$5 nm) to $\sim$1000 cm$^2$V$^{-1}$s$^{-1}$ (at a thickness of $\sim$10 nm).\cite{ref15,ref16} Nevertheless, new 2D layered materials with moderate direct bandgap and high carrier mobility are still highly sought. In this work, we show an ab initio calculation evidence of a new 2D layered material -- the TiS$_3$ monolayer sheet -- with the desired electronic properties.

Historically, bulk materials such as graphite, TMDCs and black phosphorous were studied well ahead of their 2D layered-material counterparts. Likewise, properties of bulk TiS$_3$ are known much earlier than those of 2D form. Bulk TiS$_3$ has a monoclinic crystalline structure (with the space group of p21/m), and the TiS$_3$ crystal can be viewed as stacked parallel sheets with each sheet being composed of 1D chains of triangular TiS$_3$ unit. These sheets interact with one another via the van der Waals (vdW) forces.\cite{ref28,ref29} It is also known that materials with stacking-layer structures can be a good precursor for contriving 2D atomic layers via either exfoliation\cite{ref30,ref31} or mechanical cleavage\cite{ref32}. Several electrical and transport measurements have been reported,\cite{ref33,ref34} showing that the bulk TiS$_3$ is an n-type semiconductor with carrier mobility of $\sim$30 cm$^2$V$^{-1}$s$^{-1}$ at room temperature. The mobility can be further enhanced up to $\sim$100 cm$^2$V$^{-1}$s$^{-1}$ at the low temperature 100 K.\cite{ref34} Moreover, optical absorption measurements indicate that the bulk TiS$_3$ exhibits an optical gap about 1 eV.\cite{ref35} More importantly, several recent experiments demonstrate that thin films of TiS$_3$ with thickness of 10$^2$ nanometers possess a direct bandgap of $\sim$1.1 eV and exhibit good photo response.\cite{ref36,ref37,ref38} The moderate bandgap of bulk TiS$_3$ coupled with relatively high carrier mobility renders the bulk TiS$_3$ a highly promising precursor for isolating 2D TiS$_3$ sheets with desired properties for nanoelectronic applications.
 \begin{figure}
 \centering
 \includegraphics[width=0.9\linewidth,clip=] {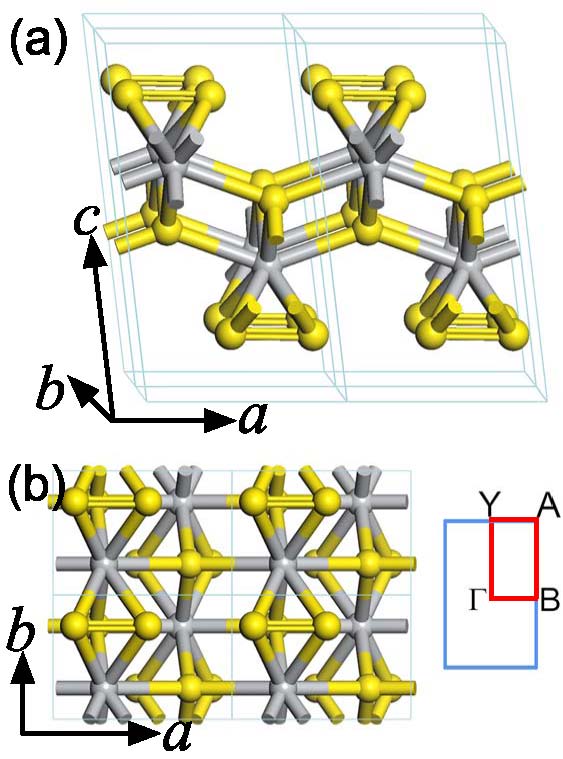}
 \caption{(Color online)(a)A 2$\times$2$\times$1 supercell of the bulk TiS$_3$ structure, (b) top view of a 2$\times$2 TiS$_3$ monolayer sheet (left), and the first Brillouin zone and the high symmetry points associated with the monolayer (right). The grey and yellow spheres refer to Ti and S atoms, respectively.}\label{fig1}
 \end{figure}
For the 2D TiS$_3$ monolayer sheet, geometrical optimization and electronic structure calculations are carried out using density-functional theory (DFT) methods within the generalized gradient approximation (GGA) and with the Perdew-Burke-Ernzerhof (PBE) exchange correlation functional, as implemented in the Vienna ab initio simulation package (VASP).\cite{ref39} The Grimme's D2 dispersion correction\cite{ref40} is adopted to account for the long-range vdW interactions.  The ion-electron interaction is treated using the projector-augment-wave (PAW) technique and a kinetic energy cutoff of 500 eV is chosen. A vacuum space of $\sim$20 \AA \ along the direction normal to the monolayer plane is undertaken so that the interlayer interaction due to the periodic boundary condition can be neglected. For the geometric optimization, a 7$\times$10$\times$1 Monkhorst-Pack\cite{ref41} grid is used and all structures are relaxed until the forces on the atoms are less than 0.01 eV/\AA, and the total energy change becomes less than 1.0 $\times$ 10$^{-5}$ eV. For total energy calculations, a fine 35$\times$50$\times$1 grid is adopted. Since the PBE functional tends to underestimate the band gap of semiconductors, the hybrid HSE06 functional\cite{ref42} is also used to compute the band gap of optimized TiS$_3$ monolayer sheet.

The carrier mobility ($\mu$) is calculated based on the deformation theory proposed by Bardeen and Shockley.\cite{ref43} Due to the fact that for inorganic semiconductors, the coherent wavelength of thermally activated electrons or holes is close to the acoustic phonon wavelength and is much longer than typical bond length, the scattering of a thermal electron or hole is dominated by the electron-acoustic phonon coupling.\cite{ref43} The deformation theory has been widely used to evaluate $\mu$ of low dimensional systems.\cite{ref19,ref44,ref45,ref46,ref47,ref48} On the basis of effective mass approximation, the charge mobility in 2D materials can be expressed as:
\begin{equation}
\mu=\frac{2e\hbar^3C}{3k_BT|m^*|^2E_1^2}
\end{equation}
Here, $C$ is the elastic modulus defined as
$C=[\partial^2E/\partial\delta^2]/S^0$,
where $E$ is the total energy of the system (per supercell), and $\delta$ is the applied uniaxial strain, and $S^0$ is the area of the optimized 2D structure. $m^*$ is the effective mass, which can be given as $m^*=\hbar^2 (\partial^2 E/\partial k^2 )^{-1}$ (where $\hbar$ is the Planck's constant and $k$ is magnitude of the wave-vector in momentum space), $T$ is the temperature, and $E_1$ is the deformation potential constant, which is proportional to the band edge shift induced by the strain. $E_1$ is defined as $\delta E=E_1\times(\delta l/l_0)$, where  $\delta E$ is the energy shift of the band edge position with respect to the lattice dilation $\delta l/l_0$ along either direction $a$ or $b$, the energies of the band edges are calculated with respect to the vacuum level.

The PBE-D2 optimized structure of bulk TiS$_3$ is shown in Fig.\ref{fig1}(a), and the associated lattice constants are $a$ = 4.982 \AA, $b$ = 3.392 \AA \ and $c$ = 8.887 \AA, and lattice angle $\beta$=97.24$^\circ$, all in very good agreement with the experimental results, $a$ = 4.958 \AA, $b$ = 3.401 \AA and $c$ = 8.778 \AA, and $\beta$=97.32$^\circ$.\cite{ref28} Furthermore, the computed band structures of the bulk TiS$_3$ from both PBE-D2 and HSE06 are shown in Supplemental Material\cite{new}. PBE-D2 and HSE06 give qualitatively the same results except for the band gap. Both PBE-D2 and HSE06 calculations indicate that the bulk TiS$_3$ is an indirect gap semiconductor from $\Gamma$ (0, 0, 0) to Z (0, 0, 0.5). The PBE-D2 computation gives a band gap of 0.21 eV, while HSE06 gives 1.02 eV. The latter agrees well with the measured optical gap which is around 1 eV.\cite{ref35} The good agreement between the benchmark calculations and experiments for the bulk TiS$_3$ show that the theoretical methods chosen for this system is reliable. In addition, the band structures near the conduction band minimum (CBM) or the valence band maximum (VBM) of the bulk TiS$_3$ exhibit notable in-plane dispersion behavior (from $\Gamma$ to Y (0, 0.5, 0)) or $\Gamma$ to B (0.5, 0, 0)), indicating that the 2D TiS$_3$ monolayer sheet may have relatively high carrier mobility.
 \begin{figure}
 \centering
 \includegraphics[width=0.9\linewidth,clip=] {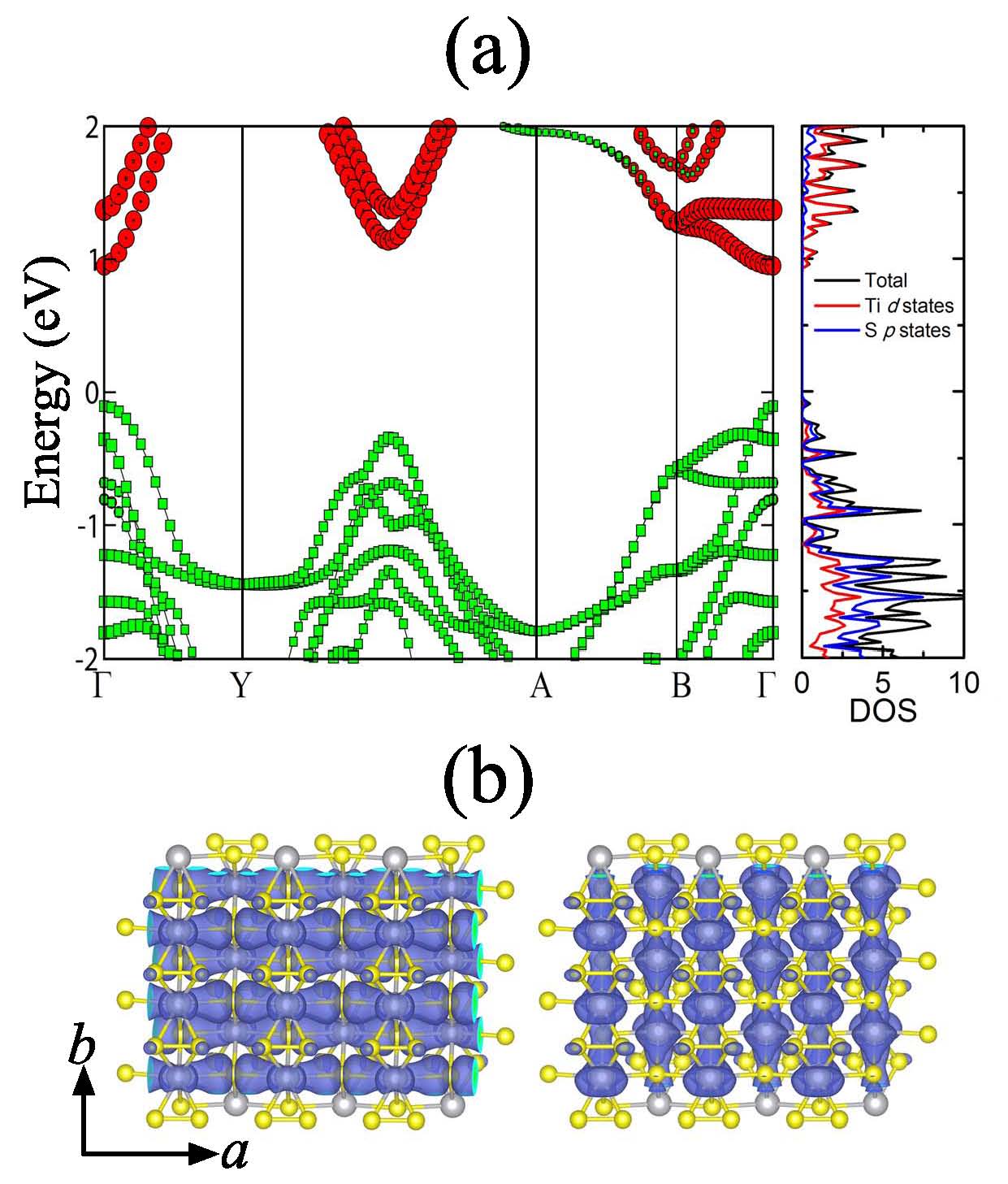}
 \caption{(Color online)(a) Computed HSE06 band structures of TiS$_3$ monolayer sheet; $\Gamma$ (0.0, 0.0, 0.0), Y (0.0, 0.5, 0.0), A (0.5, -0.5, 0.0), B (0.5, 0.0, 0.0) refer to the special points in the first Brillouin zone; red circles, green rectangles refer to the contributions of Ti $d$ states and S $p$ states; and the Fermi level is set to zero; (b) iso-surface plots of the charge density of VBM (left) and CBM (right) of the TiS$_3$ monolayer sheet, with an iso-value of 0.003 $e$/Bohr$^3$. }\label{fig2}
 \end{figure}
 The computed HSE06 band structures and density of states (DOS) of the TiS$3$ monolayer are shown in Fig.\ref{fig2}(a). Since the original Z point of the bulk TiS$_3$ folds back to the $\Gamma$ point for the TiS$_3$ monolayer, the TiS$_3$ undergoes an indirect-direct transformation from an indirect band gap semiconductor for the bulk to a direct band gap semiconductor for the 2D monolayer counterpart, akin to the case of MoS$_2$.\cite{ref18} Both VBM and CBM are located at the $\Gamma$ point, yielding a direct band gap of 1.06 eV. Moreover, from the orbital and atom projected DOS, we can see that the valance bands exhibit strong hybridization between the S $p$ states and Ti $d$ states from -2 eV to the top of valence band, while the conduction bands are mainly contributed from the d states of Ti (see Fig.\ref{fig2}(a)). The isosurface plots of the VBM and CBM are shown in Fig.\ref{fig2} (b), which show that the holes (from VBM) favor the $a$ direction, while the electrons (from CBM) favor the $b$ direction.
 \begin{figure}
 \centering
 \includegraphics[width=0.9\linewidth,clip=] {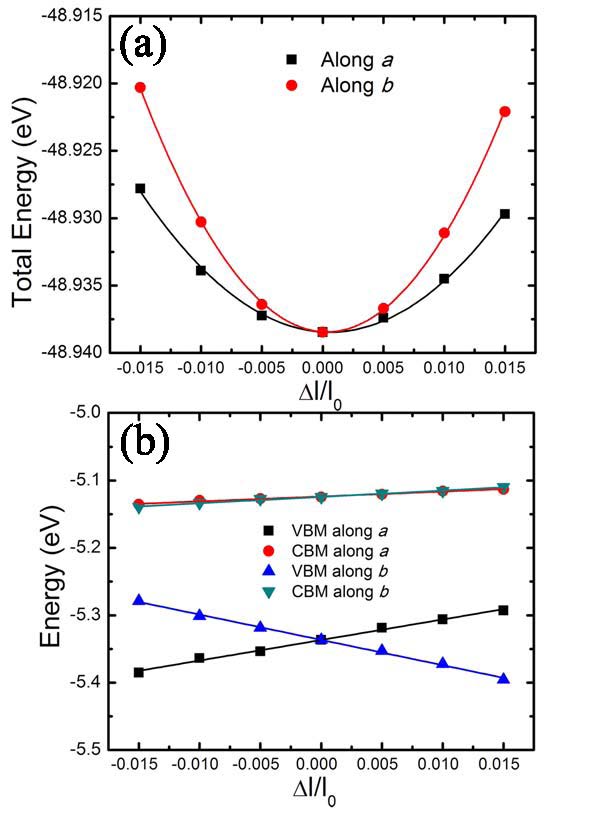}
 \caption{(Color online)(a) Strain-total energy relations and (b) shifts of VBM and CBM under uniaxial strain along $a$ and $b$ directions for TiS$_3$ monolayer sheet, $\delta l$ refers to the dilation along $a$ or $b$, while $l_0$ refers to the lattice constant of $a$ or $b$ at equilibrium geometry. In (b), the vacuum level is set at zero for reference. }\label{fig3}
 \end{figure}
 To compute the 2D elastic modulus ($C$) and the deformation-potential constant ($E_1$), we dilate the lattice of the cell up to 1.5\% along both $a$ and $b$ directions, and then calculate the total energy and the positions of CBM and VBM with respect to the dilation. The atomic positions are relaxed at the dilation, and the electronic energies are calculated at the PBE-D2 level with the ultra-fine k-meshes (35$\times$50$\times$1). We note that although the PBE functional underestimates the band gap, it can give quite good carrier mobility data for MoS$_2$,\cite{ref44} graphene,\cite{ref46} graphyne\cite{ref45} and graphdiyne\cite{ref47}. The total energy-strain relation and the positions of CBM and VBM with respect to the strain are plotted in Fig.\ref{fig3}. As shown in Fig.\ref{fig3}(b), the response of CBM and VBM to the applied strain appears to be highly anisotropic. The CBM increases monotonously with the strain either along $a$ or $b$ direction, while the VBM decreases monotonously with the strain along the $b$ direction but increases along $a$, resulting in band gap increase due to the strain along $b$ but band gap decrease along $a$. The 2D modulus ($C$) is attained by the quadratic fitting of the total energy versus strain, and the deformation potential constant ($E_1$) is calculated by the linear fitting of the CBM (VBM)-strain relation. With $C$, $E_1$ and the effective mass known, the carrier mobilities are calculated via equation (1). These data and the relaxation time ($\tau=\mu m^*/e$) are summarized in Table I.
\begin{table}
\caption{Calculated deformation-potential constant ($E_1$), 2D modulus ($C$), effective mass ($m^*$), relaxation time ($\tau$), and electron and hole mobility ($\mu$) in $a$ and $b$ directions of TiS$_3$ monolayer sheet at 300 K.}
\begin{tabular}{|c|c|c|c|c|c|}
\hline
\hline
 & $E_1$ (eV) & $C$ (N/m)& $m^*$ (m$_e$)& $\tau$ (ps) &  $\mu$ (10$^3$ cm$^2$V$^{-1}$s$^{-1}$) \\
\hline
electron ($a$) & 0.73 & 81.29 & 1.47 & 0.84 & 1.01 \\
hole ($a$) & 3.05 & 81.29 & 0.32 & 0.22 & 1.21 \\
electron ($b$) & 0.94 & 145.05 & 0.41 & 3.23 & 13.87 \\
hole ($b$) & -3.76 & 145.05 & 0.98 & 0.085 & 0.15 \\
\hline
\hline
\end{tabular}
\label{table1}
\end{table}

As shown in Table\ref{table1}, the 2D modulus along $b$ is nearly two times higher than that along $a$ direction. This is because the Ti-S bond strength along $a$ is weaker than that along $b$ direction as the Ti-S bond length is 2.65 \AA \ along $a$ while 2.45 \AA \ along $b$. The difference between the Ti-S bond strength along $a$ and $b$ also makes the deformation-potential constant along $b$ larger than that along $a$, as the band energies are more sensitive to dilations along $b$ than along $a$. The effective mass also shows an anisotropic feature: in $a$ direction, the effective mass of hole is much smaller than that of electron, while in $b$ direction, the effective mass of electron is about half of that of the hole. These results can be well explained by a charge-density plot of CBM and VBM as shown in Figure 2(b) where one can see that VBM electrons are quite localized along $b$ while the CBM ones are delocalized along $b$ but localized along $a$.

The predicted carrier mobilities of TiS$_3$ monolayer are highly anisotropic. The computed electron mobility along $b$ direction is 13.87$\times$10$^3$ cm$^2$V$^{-1}$s$^{-1}$, about 14 times higher than that along $a$ direction (1.01$\times$10$^3$ cm$^2$V$^{-1}$s$^{-1}$), while the hole mobility along $a$ direction is 1.21$\times$10$^3$ cm$^2$V$^{-1}$s$^{-1}$, about 8 times higher than that along $b$ direction (0.15$\times$10$^3$ cm$^2$V$^{-1}$s$^{-1}$). It is worthy of mentioning that the predicted carrier mobilities are notably higher than those of the MoS$_2$ monolayer sheet (which are in the range of 60-200 cm$^2$V$^{-1}$s$^{-1}$)\cite{ref44}. Especially, along the $b$ direction, the electron mobility is about 100 times higher than the hole mobility, making $b$ direction more favorable for the electron conduction. The large difference in electron/hole mobility can be exploited for electron/hole separation.

\begin{figure}
 \centering
 \includegraphics[width=0.9\linewidth,clip=] {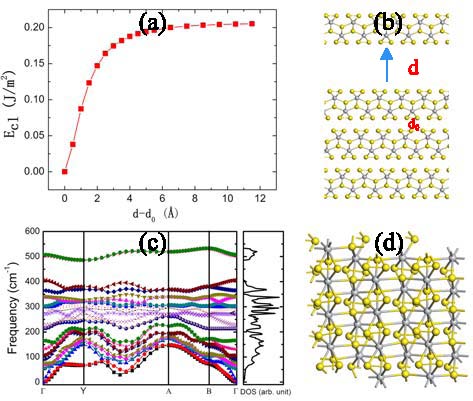}
 \caption{(Color online)(a) Cleavage energy E$_{cl}$ as a function of the separation between two TiS$_3$ monolayers. (b) Schematic view of the exfoliation of a TiS$_3$ monolayer from bulk, (c) Phonon band structure and density of states of TiS$_3$ monolayer and (d) Snapshot of TiS$_3$ monolayer at 8 ps of the BOMD simulation in the $NPT$ ensemble at 300 K and 0 GPa.}\label{fig4}
 \end{figure}

Although TiS$_3$ monolayer exhibits some novel properties for potential nanoelectronic applications, feasibility of isolation of TiS$_3$ monolayer sheet via either exfoliation or mechanical cleavage technique has yet to be confirmed. To example this feasibility, we calculate the cleavage energy by introducing a fracture in the bulk TiS$_3$ (see Fig.\ref{fig4}(b)). To this end, the total energies under variation of the separation d between the fractured parts are computed to simulate the exfoliation process.\cite{ref49,ref50} The resulting cleavage energy is plotted in Fig.\ref{fig4} (a). It can be seen that the total energy increases with the separation $d$ and gradually converges to the ideal cleavage cohesion energy of about 0.20 J/m$^2$. The latter value is less than the experimentally estimated cleavage energy for the graphite ( $\sim$0.36 J/m$^2$)\cite{ref51}, indicating that the exfoliation of bulk TiS$_3$ should be highly feasible experimentally. The stability of TiS$_3$ monolayer is another issue that should be examined. First, we compute the phonon spectrum of the TiS$_3$ monolayer, based on density functional perturbation theory with the linear response as implemented in the QUANTUM-ESPRESSO package.\cite{ref52} As shown in Fig.\ref{fig4} (c), the TiS$_3$ shows no imaginary phonon mode, indicating its dynamical stability. Next, we perform BOMD simulations. The constant-temperature (300 K) and -pressure (0 GPa) ($NPT$) ensemble is adopted. Here, the time step is 2 fs and the total simulation time is 8 ps. As shown in Fig.\ref{fig4} (d), the in-plane structure integrity of TiS$_3$ monolayer is well kept during the BOMD run, suggesting good thermal stability of the TiS$_3$ monolayer.

 In conclusion, we predict a new 2D material, TiS$_3$ monolayer sheet, which is a semiconductor with a desired direct band gap of $\sim$1 eV. The electron mobilities of the TiS$_3$ monolayer are dominant and highly anisotropic. More specifically, the electron mobility along $b$ direction exhibits a very high value of 13.87$\times$10$^3$ cm$^2$V$^{-1}$s$^{-1}$, rendering the TiS$_3$ monolayer particularly attractive for future applications in nanoelectronics as its mobility is even notably higher than that of the MoS$_2$ monolayer. The computed ideal cleavage cohesion energy for TiS$_3$ is about 0.20 J/m$^2$, less than that of graphite ($\sim$0.36 J/m$^2$), indicating the isolation of a 2D TiS$_3$ monolayer can be technically attainable via either liquid exfoliation or mechanical cleavage as done for isolation of 2D graphene or MoS$_2$ sheet.\cite{ref30,ref31,ref32} Lastly, dynamic and thermal stability of TiS$_3$ is confirmed by both phonon spectrum and BOMD simulations. Thus, we expect that fabrication of 2D TiS$_3$ monolayer and measurement of its electronic properties will be likely accomplished in the near future.

This work was supported by the National Science Foundation (NSF) through the Nebraska Materials Research Science and Engineering Center (MRSEC) (grant No. DMR-1420645)NSF, UNL Nebraska Center for Energy Sciences Research, and UNL Holland Computing Center.

\end{document}